\documentclass[11pt,bibliography=totoc]{scrartcl}

\usepackage{soul}
\usepackage[utf8]{inputenc}
\usepackage[english]{babel}
\usepackage{graphicx,amsmath,amsfonts,amssymb,dcolumn,amsthm,slashed,bbm,xspace,pgffor,tikz,mathbbol,latexsym,cite,braket}
\usepackage{microtype}
\usepackage{mathtools}
\usepackage{nccmath}
\usepackage{lmodern}
\usepackage[normalem]{ulem}
\usepackage{xcolor}
\usepackage[colorlinks=true,citecolor=blue,urlcolor=blue,linkcolor=blue]{hyperref}
\usepackage[hang]{footmisc} %removes the footer margin space
\usepackage[affil-it]{authblk}
\usepackage{multicol}

\numberwithin{equation}{section}

\begin{document}

\title{\textbf{Holographic QCD phase diagram for a rotating plasma in the Hawking-Page approach
}}

\author{Nelson R.~F.~Braga$^{a}$\thanks{\href{mailto:braga@if.ufrj.br}{ braga@if.ufrj.br}} , ~  Octavio C.~Junqueira$^{a,b}$\thanks{\href{mailto:octavioj@pos.if.ufrj.br}{octavioj@pos.if.ufrj.br}} }
\affil{ $^{a}$ UFRJ --- Universidade Federal do Rio de Janeiro, Instituto de Física,\\
Caixa Postal 68528, Rio de Janeiro, Brasil
\vspace{0.2cm}

$^{b}$ UFABC --- Universidade Federal do ABC, Centro de Matemática,\\
Computação e Cognição, 09210-580, Santo André, Brasil}

\date{}
\maketitle

\begin{abstract}
We investigate the combined effect of rotation and finite chemical potential in the confinement/deconfinement transition of strongly interacting matter. 
The holographic description consists of a five-dimensional geometry that contains a black hole (BH) in the deconfined (plasma) phase.  The geometry is equipped with some cut-off that introduces an infrared energy scale. We consider two possibilities: the so-called hard wall and soft wall AdS/QCD models. The transition between the plasma and hadronic phases is represented holographically as a Hawking-Page transition between geometries with and without a black hole. The gravitational dual of the rotating plasma at finite density is given by a Reissner-Nordström (RN) charged anti-de Sitter (AdS) BH with non-zero angular momentum. 
This analysis provides the critical temperature of deconfinement as a function of the quark chemical potential and the plasma rotational velocity. For the case of very low temperatures, the dependence of the critical values of the chemical potential for the transition to occur at $ T \to 0 $ on the rotation is found.

\end{abstract}

\section{Introduction}

The Quark-Gluon Plasma (QGP)  is formed experimentally through ultra-relativistic collisions of heavy nuclei, produced in  particle accelerators. This state of matter is composed of deconfined partons that interact strongly 
\cite{Shuryak:2008eq,Casalderrey-Solana:2011dxg,Busza:2018rrf}. For non-central heavy-ion collisions, the resulting plasma acquires a large angular momentum, of the order of $10^3\, \hbar$, with angular velocity $\omega \sim 0.03 \, fm^{-1}c$. This estimate was obtained in 2017, when the global hyperon polarization was observed in $Au+Au$ collisions at Relativistic Heavy Ion Collider (RHIC) \cite{STAR:2017ckg}. Recent models predict that it could reach even higher values, with $\omega \sim 0.1\, fm^{-1}c$ \cite{Deng:2016gyh, Jiang:2016woz}. The measurements of the alignment between the angular momentum of a non-central collision and the spin of emitted particles reveal that the produced fluid possesses approximately the vortical structure of a liquid with almost zero viscosity \cite{Heinz:2013th}, in accordance with hidrodynamic predictions for QCD matter at high temperatures and densities. Such a result motivated a great interest in  the study of strong interactions in rotating systems. Holographic AdS/QCD models provide a powerful framework for analyzing how rotation affects the confinement/deconfinement phase transition. 

The influence of rotation on the QCD phase diagram is a topic under active investigations. Some recent studies can be found, for example, in Refs.  \cite{Miranda:2014vaa,Jiang:2016wvv,McInnes:2016dwk,Mamani:2018qzl,Wang:2018sur,Chernodub:2020qah,Arefeva:2020jvo,Chen:2020ath,Zhou:2021sdy,Braguta:2021jgn,Braguta:2021ucr,Golubtsova:2021agl,Fujimoto:2021xix,Braga:2022yfe,Chen:2022mhf,Golubtsova:2022ldm,Chernodub:2022,Zhao:2022uxc,Braguta:2023yjn, Azuma:2024bbi}. Using AdS/QCD approaches, it was found in Refs.  \cite{Chen:2020ath,Braga:2022yfe}, that the critical temperature of deconfinement decreases with increasing angular velocity. This prediction is in agreement with other phenomenological models such as the Nambu-Jona Lasinio (NJL) \cite{Wang:2018sur}. In contrast, relativistic simulations of the effect of rotation on the confinement/deconfinement phase transition in gluodynamics lattice \cite{Braguta:2021jgn,Braguta:2021ucr} indicate that the critical temperature increases with increasing angular velocity. In a recent work \cite{Zhao:2022uxc} the holographic approach was used to study the rotational effects
in the $2+1$ flavor QCD matter, based on the 5-dimensional EMD (Einstein-Maxwell-dilaton) gravitational model. The critical temperature was obtained from the computation of Polyakov loops.

Our purpose here is to use the Hawking-Page (HP) approach, considering the hard wall 
\cite{Polchinski:2001tt,BoschiFilho:2002ta,BoschiFilho:2002vd} and soft-wall \cite{Karch:2006pv} AdS/QCD models.   In order to describe the effects of plasma rotation and density we consider the holographically dual geometry containing a rotating charged AdS black hole (BH), where the BH charge is associated with the quark chemical potential in the dual gauge theory. In the  HP  approach the confinement/deconfinement transition is translated into the transition between two different geometries that are asymptotically AdS: one contains the black hole and the other is just the thermal AdS space.  This approach makes it possible to compute the combined effects of quark density and plasma rotation on the phase transition.  

This paper is organized as follows: in  Section 2, we present the rotating charged AdS black hole geometry, and discuss the relation between this 5-d gravitational background and the dual QGP. In Section 3, we apply the Hawking-Page method to the hard wall and soft wall AdS/QCD models at finite density, for rotating QCD matter. The critical temperatures of deconfinement are obtained as functions of the quark chemical potential and the rotation velocity. The QCD phase diagram for each model is also presented. In Section 4, we analyze the results obtained and present some conclusions.

\section{Rotating charged black hole in AdS space}
 
In the holographic approach to QCD, the four dimensional gauge theory is assumed to be dual to a five dimensional geometry that is asymptotically anti-de Sitter ($AdS_5$). Such space is a 
solution of Einstein's equations with a negative cosmological constant $ \Lambda = - \frac{12}{L^2} $,  and constant curvature $R = - \frac{20}{L^2}  $, where $L$ is the radius of the space. 
In particular, for a dual description of a gauge system with finite temperature and non-zero quark chemical potential, the gravitational dual is a charged AdS black hole space. 

For the Reissner-Nordström (RN) charged AdS BH in the Euclidean
signature with compact time direction, the metric is given by \cite{Lee:2009bya}
\begin{equation}\label{BHAdS}
ds^2= \frac{L^2}{z^2}\left( f(z,q) dt^2 + d\overrightarrow{x}^2 + \frac{dz^2}{f(z,q)} \right)\;,
\end{equation}
with 
\begin{equation}\label{f(z)}
f(z,q) = 1 - \frac{z^4}{z_h^4}-q^2 z_h^2 z^4 + q^2 z^6\;,
\end{equation}
where $z_h$ is the location of the horizon, such that $f(z_h,q) = 0$, and $q$ is a parameter that is proportional to the BH charge. In AdS/QCD, this metric corresponds to the deconfined plasma phase. On the other hand, the hadronic phase is described by the thermal AdS space with line element
\begin{equation}\label{ThAdS}
ds^2= \frac{L^2}{z^2}\left( dt^2 + d\overrightarrow{x}^2 + dz^2 \right)\;,
\end{equation}
which is equivalent to Eq. \eqref{BHAdS} taking $f(z,q) = 1$. Without rotation, the compactified time  coordinate is periodic with the BH period given by $\beta = 1/T$, being $T$ the BH temperature. This temperature is associated with the BH charge parameter and its horizon position, according to the equation
\begin{equation} \label{TH}
T(q) = \frac{\vert \partial_zf(z,q)\vert_{(z = z_h)}}{4\pi} = \frac{1}{\pi z_h} \left(1-\frac{q^2 z_h^6}{2}\right) \;,
\end{equation}
that comes from the condition of absence of conical singularity at the horizon \cite{Hawking:1982dh}.

For a system with quarks, the energy change caused by a variation  $ \delta  J^0 $ of the density $ J^0 \equiv \bar{\psi} \gamma^0 \psi$ is proportional to the chemical potential ($\mu$): $ \mu \, \delta  J^0 $. 
In the holographic description the chemical potential works as the source of the correlation functions of the gauge theory density operator. The Reissner-Nordström solution includes not only the metric \eqref{BHAdS} but also an Abelian field $V_\mu $ living in the five dimensional space and associated with the black hole charge. The time component $V_0$ will couple to the density $ J^0 $. The boundary value of $V_0$ acts as the source for the density correlators and thus can be associated with $\mu$.  One can choose a particular RN solution for the vector field with all components vanishing, except the temporal one. For small values of $z$ \cite{Colangelo:2010pe, Lee:2009bya} this solution reads, after a Wick rotation,
\begin{eqnarray}
V_0 &=& A_0(z) \;= i(C - Qz^2)\;,\nonumber \\
V_i &=& V_z = 0 \quad (i = 1,2,3)\;,\label{V} 
\end{eqnarray} 
where $C$ is a constant, and $Q$ is the black hole charge density, as can be shown by using the Gauss law. This charge density, that we will call simply as charge for simplicity,  is related to the RN metric parameter $q$ through (see \cite{Lee:2009bya}):
\begin{equation}\label{Q}
Q^2 = \frac{3 g_5^2 L^2}{2 \kappa^2}q^2\;,
\end{equation}
where $\kappa^2$ and $g_5^2$  are the gravitational and gauge couplings, respectively, that appear in the Einstein-Hilbert action with the gauge field, which will be introduced in Section 3.

From this interpretation, in Euclidean space, $A_0(0) = i \mu$. Using this in Eq. \eqref{V} one finds $C = \mu$ and
\begin{equation} \label{A0}
    A_0(z) = i(\mu - \eta q z^2)\;,
\end{equation}
where
 \begin{equation} \label{eta}
\eta = \frac{Q}{q} = \sqrt{\frac{3 g^2_5 L^2}{2 \kappa^2}} .
\end{equation}
 
By imposing the Dirichlet boundary condition $A_0(z_h) = 0$, which is required to obtain a gauge field with regular norm, see \cite{Horigome:2006xu, Nakamura:2006xk, Hawking:1995ap, Ballon-Bayona:2020xls}, we find the following relation between the charge $q$ and the quark chemical potential:
\begin{eqnarray}\label{muq}
    \mu = \eta q z_h^2 = Q z_h^2\;,
\end{eqnarray}
that will be used to describe the QCD phase diagram.

In order to introduce rotation in the system, one has to add angular momentum to the black hole. In this work, we will assume a plasma with cylindrical symmetry,  which is obtained by considering the case when the  BH metric \eqref{BHAdS}, take the form
\begin{equation}\label{CBH}
	ds^2 = \frac{L^2}{z^2} \left( f(z,q)dt^2 + l^2 d\phi^2 + \sum_{i=1}^2 dx_i^2 + \frac{dz^2}{f(z,q)}\right)\;, 
\end{equation}
where  $l$ is  the radius of a hyper-cylinder and $ 0 \le \phi \le 2 \pi $. The BH angular momentum can be added by performing the coordinate transformation \cite{BravoGaete:2017dso, PhysRevD.97.024034}  
\begin{eqnarray}
	t &\rightarrow& \frac{1}{\sqrt{1- \omega^2 l^2}} \left(t + l^2 \omega \phi \right)\;,\label{T1}\\
	\phi &\rightarrow& \frac{1}{\sqrt{1- \omega^2 l^2}} \left(\phi +  \omega t \right)\label{T2}\;,
\end{eqnarray}
that corresponds to a change to an observer for which the angular coordinate is varying uniformly with time, with   angular velocity $\omega$. 

The coordinate transformations \eqref{T1} and \eqref{T2} act both on the metric and on the gauge field. Using Eq. \eqref{A0}, the transformed gauge field is given by
\begin{eqnarray}
    A^\prime_0 &=& \gamma(\omega l) A_0 = i( \gamma \mu - \eta \gamma q z^2)\;,\label{A0prime}\\
    A_\phi^\prime & = &-l^2\omega \gamma(\omega l) A_0\label{Aphiprime}\\
    A_{x_1}^\prime &=& A_{x_2}^\prime = A^\prime_z = 0\;, \label{AiAzprime}
\end{eqnarray}
where $\gamma(\omega l ) = \frac{1}{\sqrt{1-\omega^2l^2}}$ is the Lorentz factor.

For the case of static plasma, described by the metric \eqref{CBH} in the presence of the vector field of eq.\eqref{A0}, the chemical potential was identified as $A_0 (z=0) $, up to an $i$-factor that comes from the Wick rotation. So, one could think that for the rotating plasma, represented by the metric obtained after the coordinate transformations \eqref{T1} and \eqref{T2} one could associate $A^\prime_0 (0) $ to the chemical potential. However, there is a subtle point. For the particular case of the static plasma, the energy density associated with the interaction of the charge density with the gauge potential is given by the single term $ A_0 J^0  $. However, in the rotating case, the current is also transformed through the transformations \eqref{T1} and \eqref{T2}:
\begin{equation}
    J^{\prime \,  0}  = \gamma(\omega l) J^0  \,\,,\, \, J^{\prime \, \phi}  = \gamma \omega J^0 = \omega  J^{\prime \, 0} \label{newJ} \,.
\end{equation}
The general form of the interaction term, that is invariant under coordinate transformation, is $ A_M J^M  $. For the rotating case one has:
\begin{equation}
 A^\prime_M    J^{\prime \,  M}  = A^\prime_0    J^{\prime \,  0} + A^\prime_\phi    J^{\prime \,  \phi } \,=\, A^\prime_0    J^{\prime \,  0} \frac{1}{\gamma^2 } \,.
  \label{newINT} 
\end{equation}
Then, a variation $ \delta J^{\prime \,  0}$ of the density of the rotating plasma leads to a variation of the interaction term: $ A^\prime_0  \,  \delta J^{\prime \,  0} \,  \frac{1}{\gamma^2 } $. Therefore, the chemical potential in the rotating plasma should be defined as: 
\begin{eqnarray}\label{muprime}
   \mu^\prime =  \frac{\mu}{\gamma}\;.
\end{eqnarray}
where $\mu^\prime$ corresponds to the quark chemical potential of the rotating plasma, while $\mu$ is the chemical potential for the plasma at rest. This relation between the transformed $\mu^\prime$ for the rotating system and $\mu$ is in agreement with the one obtained in \cite{Chen:2020ath,Zhao:2022uxc}.

Replacing \eqref{T1} and \eqref{T2} in Eq. \eqref{CBH}, one obtains the following metric for the rotating BH in the canonical form \cite{Zhou:2021sdy, LEMOS199546, BravoGaete:2017dso}
\begin{eqnarray}\label{canon}
	ds^2 = N(z,q) dt^2 + \frac{L^2}{z^2}\frac{dz^2}{f(z,q)} + R(z,q)\left( d\phi + P(z,q) dt\right)^2 + \frac{L^2}{z^2} \sum_{i-1}^2 dx_i^2\;,
\end{eqnarray}
with 
\begin{eqnarray}
	N(z,q) &=& \frac{L^2}{z^2} \frac{ f(z,q) (1-\omega^2 l^2)}{1- f(z,q)\omega^2 l^2}\;, \\
	R(z,q) &=& \frac{L^2}{z^2}\left( \gamma^2 l^2 -  f(z,q) \gamma^2 \omega^2 l^4\right)\;, \\
	P(z,q) &=& \frac{\omega(1-f(z,q))}{1- f(z,q)\omega^2 l^2}\;. 
\end{eqnarray}

The black hole temperature is obtained from the surface gravity formula. Defining $h_{00}(z) \equiv - N(z)$, see \cite{Zhou:2021sdy}, one has
\begin{eqnarray}\label{HTrot}
	T(q, \omega) &=& \vert \frac{\kappa_G}{2\pi} \vert = \bigg\vert \frac{\lim_{z\rightarrow z_h}- \frac{1}{2} \sqrt{\frac{g^{zz}}{-h_{00}(z)}}h_{00,z}}{2\pi}\bigg \vert = \frac{1}{\pi z_h} \left(1-\frac{{q}^2 z_h^6}{2}\right) \sqrt{1-\omega^2 l^2}\;, 
    \end{eqnarray}
where $\kappa_G$ is the surface gravity, and $g^{zz}$ the $z-z$ component of the inverse of the cylindrical BH metric. One observes that $T(q,\omega) =  \frac{ T(q,0)}{\gamma(\omega l)}$. Equation \eqref{HTrot} implies a physical condition for the positivity of the temperature:  
\begin{equation}\label{positivity}
  0 \leq z_h \leq \left(\sqrt{2}/q\right)^{1/3}\;.
\end{equation}
 Note that in the limit of $q = 0$, the temperature goes to zero when $z_h \rightarrow \infty$, and goes to infinity when $z_h \rightarrow 0$.  The map between $T$ and $z_h$ is modified at finite density.  For $z_h \rightarrow \left(\sqrt{2}/q\right)^{1/3}$, $T \rightarrow 0$, while $T \rightarrow \infty$ when $z_h \rightarrow 0$. So, the holographic model describes all possible plasma temperatures, with $z_h $ in the interval \eqref{positivity}. 

 By requiring that the asymptotic limits of the thermal and BH AdS geometries in the rotating system are the same at $z=\epsilon$, with $ \epsilon \to 0 $, one finds that the thermal AdS period is
\begin{equation}\label{betaAdS}
\beta_{AdS} (q,\omega) = \beta(q, \omega) \sqrt{f(\epsilon, q)}\;,
\end{equation}
where $\beta(q, \omega) = 1/T(q, \omega)$, as defined by Eq. \eqref{HTrot}. 
In the next section we compute the regularized black hole actions in the hard and soft wall AdS/QCD models and use the Hawking-Page approach in order to analyze the effects of rotation and density in the confinement/deconfinement transition.

\section{Hawking-Page transition and QCD phase diagram}
In the hard \cite{Polchinski:2001tt,BoschiFilho:2002ta,BoschiFilho:2002vd} and soft  \cite{Karch:2006pv} wall holographic AdS/QCD models, one introduces an energy parameter in the AdS geometry, that is interpreted as an infrared (IR) cutoff in the gauge theory side of the duality. In the first case, this is done by imposing that the $z$ coordinate has a maximum value: $ 0 \le z \le z_0$. In the second case, one introduces in the geometry a dilaton background $\Phi (z)$ containing a mass scale that breaks conformal symmetry.

In Euclidean space, one can write the five-dimensional  gravitational action for both models  in the general form\cite{Herzog:2006ra,BallonBayona:2007vp}
\begin{equation}\label{action1}
	I_G = - \frac{1}{ 2 \kappa^2} \int dz\int d^4x \sqrt{g} e^{-\Phi}\left( R - \Lambda \right) \;
\end{equation}
where $\Phi(z) = cz^2$, and $\kappa$ is the gravitational coupling associated with the Newton constant. For the hard wall model one chooses $ c = 0$ and considers $z \leq z_0$, where $1/z_0$ is the IR energy parameter. For the soft wall, $z_0 \to \infty $ and $ \sqrt{c} $ is the energy parameter. 
For the rotating charged AdS BH and thermal AdS spaces the determinant of the metric is the the same, given by $g = \frac{L^{10}}{z^{10}}$, such that the finite temperature version of the gravitational on-shell action \eqref{action1} reads
\begin{equation}
	I_{G_{on-shell}} = \frac{4L^3}{\kappa^2} V_{3D} \int_0^{\beta_s} dt \int_0^{z_{min}} dz\, z^{-5}e^{-cz^2}\;\,,
\end{equation}   
    where $z_{min}$ is the minimum of $( z_0 , z_h )$, and $\beta_s$ is the period of the Euclidean time coordinate in the corresponding space. This expression was obtained using the AdS curvature and cosmological constant presented in section \textbf{2}. The thermal AdS space has no horizon, or equivalently $z_h \rightarrow \infty$. Integration over spatial bulk coordinates $x$ is trivial and generates the volume factor $V_{3D}$. 

The action for the vector fields is 
\begin{equation}\label{actionVF}
    I_{VF} = -\frac{1}{4g_5^2}\int_0^{z_{min}} dz \int d^4x \sqrt{g} e^{-\Phi} F_{MN}F^{MN}\;,
\end{equation}
where $F_{MN} = \partial_{M} V_N - \partial_N V_M$. As discussed previously, see Eq. \eqref{V}, the boundary value of $A_0 $ works as the source for the quark density. Thus, replacing the transformed vector field given by Equations \eqref{A0prime}, \eqref{Aphiprime} and \eqref{AiAzprime}, and the rotating metric \eqref{canon} into Eq. \eqref{actionVF}, one gets the on-shell $U(1)$ gauge action
\begin{equation}\label{action2}
    I_{VF_{on-shell}} =  \frac{2L \eta^2 q^2}{g^2_5 }\gamma^4 V_{3D} \, \int_0^{\beta_s} dt  \int_\epsilon^{z_{min}}dz\, z e^{-c z^2}\left[(1+l^2\omega^2)^2 + 4l^2\omega^2f(z,q)\right]\;.
\end{equation}
with the time coordinate varying from $0$ to $\beta(q,\omega)$ for the BH geometry, and from $0$ to $\beta_{AdS}(q,\omega)$ for the thermal AdS one. The total on-shell action for the rotating QCD matter at finite density is then  
\begin{equation}\label{totalI}
    I_{on-shell} = I_{G_{on-shell}} + I_{VF_{on-shell}}\;. 
\end{equation}
 As the trivial volume factor of the bulk appears in both actions, $I_G$ and $I_{VF}$, we define the action density dividing the action  by $V_{3D}$, $\mathcal{E} = \frac{1}{V_{3D}} I_{on-shell}$. The result is
\begin{equation}
	\mathcal{E}_s(\varepsilon) =  \beta_s  \int_\varepsilon^{z_{min}}dz\, \frac{e^{-c z^2}}{z^5}\left[ \frac{4L^3}{\kappa^2} + \frac{2L\eta^2q^2}{g^2_5}\gamma^4\left((1+l^2\omega^2)^2+4l^2\omega^2f(z,q)\right) z^6\right]\;, 
\end{equation}
where we introduced the ultraviolet (UV) regulator $\varepsilon$ in the integration over $z$. The action densities of both spaces are infinite in the limit $\varepsilon \rightarrow 0$. One can eliminate this singularity by defining a regularized black hole action density as the difference between the action densities of the two geometries,  \begin{equation}\label{DeltaE}
	\bigtriangleup \mathcal{E}(\varepsilon) = \lim_{\varepsilon \rightarrow 0} \left[\mathcal{E}_{BH}(\varepsilon) - \mathcal{E}_{AdS}(\varepsilon) \right]\;, 
\end{equation}
with
\begin{eqnarray}
	\mathcal{E}_{BH}(\varepsilon) &=& \beta\int_\varepsilon^{z_{min}}dz \frac{e^{-c z^2}}{z^5}\left[ \frac{4 L^3}{\kappa^2} + \frac{2L\eta^2q^2}{g^2_5}\gamma^4\left((1+l^2\omega^2)^2+4l^2\omega^2f(z,q)\right) z^6\right]  \;, \label{DeltaEBH}\\
	\mathcal{E}_{AdS}(\varepsilon) &=&  \, \beta \sqrt{1 - \frac{\varepsilon^4}{z^4} - q^2 z_h^2 \varepsilon^4 + q^2\varepsilon^6}\int_\varepsilon^{z_0}dz \frac{e^{-c z^2}}{z^5}\left[ \frac{4L^3}{\kappa^2} + \frac{2L\eta^2q^2}{g^2_5}\gamma^4\left((1+l^2\omega^2)^2+4l^2\omega^2f(z,q)\right) z^6\right]\;, \nonumber\\   \label{DeltaEAdS}
\end{eqnarray}
where we used Eq. \eqref{betaAdS}.  

From the regularized action density \eqref{DeltaE}, one can compute the effects of rotation and quark density to the critical temperatures of confinement/deconfinement transition. This density determines the Hawking-Page transition. When $\bigtriangleup \mathcal{E}$ is positive (negative), the BH is unstable (stable), since the free energy density of the AdS space is smaller (greater) than the black hole one. 
Using gauge/gravity duality, the thermal AdS space corresponds to the confined hadronic phase while the BH phase describes the plasma. The transition occurs when $\bigtriangleup \mathcal{E}$ vanishes. In the following subsections, we analyze the cases of hard wall \cite{Polchinski:2001tt,BoschiFilho:2002ta,BoschiFilho:2002vd}  and soft wall \cite{Karch:2006pv} AdS/QCD models.

\subsection{Hard wall model}
There is no dilaton field in the background of the hard wall model, so $c=0$ in Equations  \eqref{DeltaEBH} and \eqref{DeltaEAdS}. The regularized action density, defined by Eq. \eqref{DeltaE}, takes the form:
\begin{equation}
	\centering
	\bigtriangleup \mathcal{E}(q,
 \omega, z_h) = \left\{\begin{aligned}
		\frac{L^3\pi z_h \gamma(\omega l)}{\kappa^2(1-\frac{q^2 z_h^6}{2})} (1+ q^2 z_h^6) \frac{1}{2z_h^4}\;, \quad  z_0&< z_h\;, \\
		\frac{L^3\pi z_h \gamma(\omega l) }{\kappa^2 (1-\frac{{q}^2 z_h^6}{2})}\left[ \frac{1}{z_0^4}- \frac{1}{2 z_h^4} \left(1-q^2 h_1(\omega,z_h,z_0) + q^4 h_2(\omega,z_h,z_0)\right)
        \right] \;, \quad z_0&> z_h \;, \\ 
	\end{aligned}\right.\label{FreeHW}
\end{equation}
where we have defined
\begin{eqnarray}
h_1(\omega,z_h,z_0) &=& z_h^6 + \gamma^4(z_h - z_0)(z_h+z_0)\left[ 3z_h^4 + 3^4\omega^4z_h^4 - 2l^2\omega^2\left(2z_0^4 + 2z_0^2z_h^2-7z_h^4 \right)\right]\;,\\
h_2(\omega,z_h,z_0) &=& l^2\omega^2 \gamma^4z_h^4\left(3z_0^8 - 4 z_0^6z_h^2+2
z_h^8\right)\;.
\end{eqnarray}

\begin{figure}[!htb]
	\centering
	\includegraphics[scale=0.64]{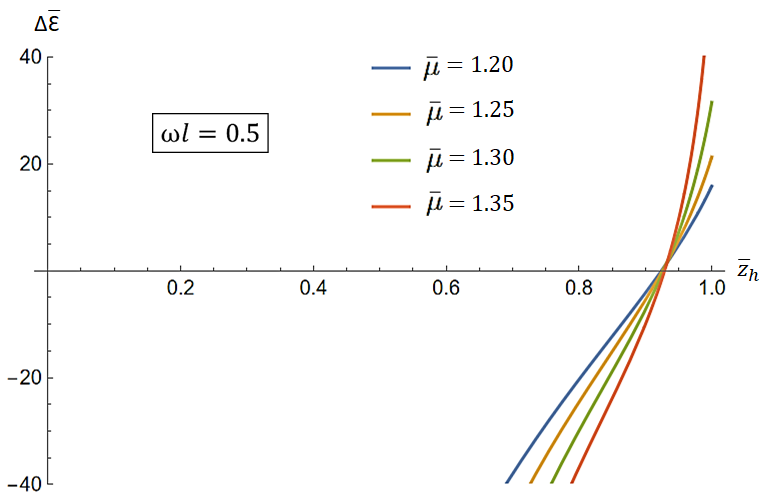}
	\caption{Regularized action density for the hard wall model at  ($\omega l = 0.5$), with  $\bar{\mu} = 1.20$ (blue), $\bar{\mu} = 1.25$ (yellow), $\bar{\mu} = 1.30$ (green), and $\bar{\mu} = 1.35$ (red). The critical horizons are defined by $\bigtriangleup \mathcal{E} = 0$.     }
\end{figure}

\begin{figure}[!htb]
	\centering
	\includegraphics[scale=0.55]{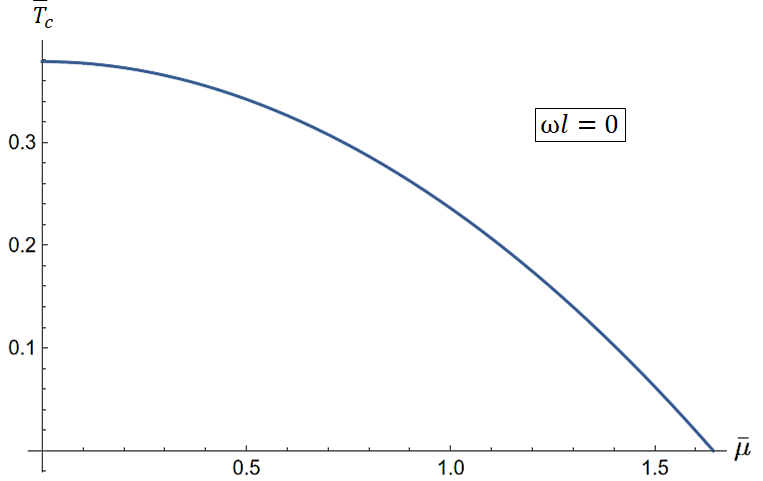}
	\caption{Critical temperature $\bar{T}_c$ versus quark chemical potential for a non-rotating plasma in the hard wall model.     }
\end{figure}

In the limit $\omega \rightarrow 0$, one recovers the result for the system at finite density without rotation of Ref.\cite{Braga:2024nnj}. Considering additionally $q \rightarrow 0$, one obtains the result for a static system at zero quark chemical potential \cite{Herzog:2006ra}. There is no transition if $z_0 < z_h$, since $\bigtriangleup \mathcal{E}$ is always positive in this case, according to the physical condition \eqref{positivity}. The critical horizon position ${z_h}_c$ is defined by the equation
\begin{equation}\label{HPtransition}
    \bigtriangleup \mathcal{E}(q,
 \omega, z_h) = 0  \quad  \text{at} \quad {z_h} = {z_h}_c(q, \omega)\;.
\end{equation}
For simplicity we will assume $\eta = 1$, so that $ \mu = q {z_h}^2$. The role of $\eta $ will be discussed in the conclusions section.  The critical temperature, as a function of $\omega l$ and $\mu$, is obtained by replacing ${z_h}_c(\mu, \omega)$ in Eq. \eqref{HTrot}.

\begin{figure}[!htb]
	\centering
	\includegraphics[scale=0.64]{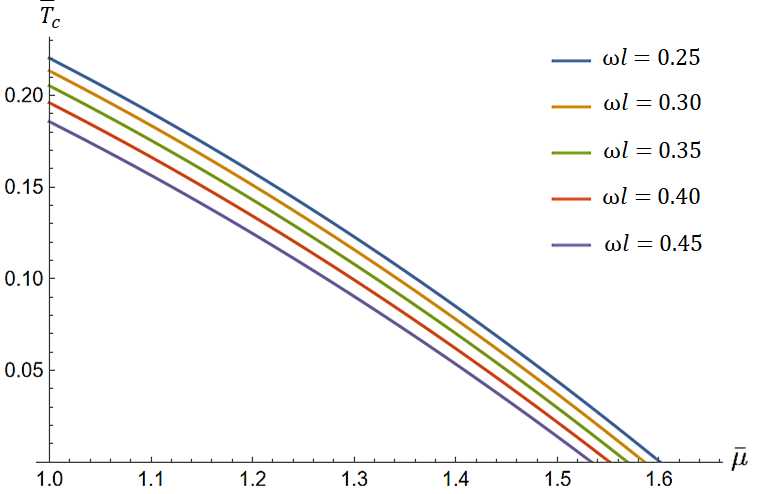}
	\caption{Critical temperatures of deconfinement as a function of quark chemical potential $\bar{\mu}$ in the hard wall AdS/QCD model, at fixed values of plasma rotational velocity.}
\end{figure}

It is interesting to work with dimensionless quantities.  Thus we define \begin{equation}\label{barDeltaE}
    \bigtriangleup \bar{\mathcal{E}}(\bar{\mu},
 \omega, \bar{z}_h) = \frac{\pi \bar{z}_h \gamma(\omega l)}{2\bar{z}_h^{12}(1-\frac{\bar{\mu}^2 \bar{z}^2_h}{2})} \left[ 2 \bar{z}_h^{12} -  \bar{z}^8_h + \bar{\mu}^2 \bar{z}_h^6 \bar{h}_1(\omega,\bar{z}_h)-\bar{\mu}^4\bar{h}_2(\omega,\bar{z}_h)\right]\,,
\end{equation}
with
\begin{eqnarray}
    \bar{h}_1(\omega,\bar{z}_h) &=& \bar{z}_h^6 + \gamma^4(\bar{z}_h - 1)(\bar{z}_h+1)\left[ 3\bar{z}_h^4 + 3^4\omega^4z_h^4 - 2l^2\omega^2\left(2 + 2\bar{z}_h^2-7\bar{z}_h^4 \right)\right]\;\\
    \bar{h}_2(\omega,\bar{z}_h) &=& l^2\omega^2 \gamma^4\bar{z}_h^4\left(3 - 4 \bar{z}_h^2+2
\bar{z}_h^8\right)\;,
\end{eqnarray}
where   $\bigtriangleup \bar{\mathcal{E}}  = \bigtriangleup {\mathcal{E}}\kappa^2 z^3_0/L^3 $ is a dimensionless version of the action density for $z_0 > z_h$, and 
\begin{eqnarray}\label{dimensionlessvar}
     \bar{z}_h &=& z_h/z_0\;, \nonumber\\
     \bar{\mu} &=&  \mu z_0\;, \nonumber\\
     \bar{q} &=& q z_0^3\;. 
 \end{eqnarray}
 
In FIG. 1, we plot $\bigtriangleup \bar{\mathcal{E}}$ as a function of $\bar{z}_h$, at fixed values of plasma rotational velocity for $z_0 > z_h$ and different values of $\bar{\mu}$. One observes that the phase transition is sensitive to variations on the quark chemical potential. The action density diverges as the system approaches the edge of the physical condition \eqref{positivity}, see Eq. \eqref{barDeltaE}. In the physical region with positive temperature, the domain of the horizon position is defined by ${z}_h \leq \sqrt{2}/\bar{\mu}$,
with $\bigtriangleup \bar{\mathcal{E}} \rightarrow \infty$ when ${z}_h \rightarrow \sqrt{2}/\bar{\mu}$. For the case with zero chemical potential, such divergencies do not appear since the pole in Eq. \eqref{barDeltaE} is not present.

For the case without rotation, the Eq. \eqref{barDeltaE} is reduced to
\begin{equation}\label{barDeltaE2}
    \bigtriangleup \bar{\mathcal{E}}(\bar{\mu},0, \bar{z}_h) = \frac{\pi \gamma(\omega l)}{2\bar{z}_h^3(1-\frac{\bar{\mu}^2 \bar{z}^2_h}{2})} \left( 2 \bar{z}_{h}^4 - 3 \bar{\mu}^{2} + 4 \, \bar{\mu}^{2} \bar{z}_{h}^2-1 \right)\,,
\end{equation}
such that the HP transition equation \eqref{HPtransition} yields the following critical horizon as
the only positive and real solution:
\begin{equation}\label{zhc}
\bar{{z_h}_c} = \sqrt{-\bar{\mu}^{ 2} + \frac{\sqrt{1 + 3 \bar{\mu}^{ 2 } + 2\bar{\mu}^{ 4} }}{\sqrt{2}}}    \,.
\end{equation}
 This equation shows that $\bar{{z_h}_c}$ increases as $\bar{\mu}$ increases. 
Replacing the critical horizon \eqref{zhc} in Eq. \eqref{HTrot}, we obtain the following expression for the dimensionless critical temperature of confinement/deconfinement for a non-rotating plasma as a function of $\mu$: 
\begin{equation}\label{barTc}
\bar{T}_c(\bar{\mu}) = \frac{1}{\pi}\left(-\bar{\mu}^{ 2}  + \frac{\sqrt{1 + 3 \bar{\mu}^{ 2} + 2\bar{\mu}^{ 4}}}{\sqrt{2}}\right)^{-1/2} \left( 1 + \bar{\mu}^{4} - \bar{\mu}^{ 2} \frac{\sqrt{1 + 3 \bar{\mu}^{ 2 }+ 2\bar{\mu}^{ 4} }}{2\sqrt{2}}  \right)\;,  
\end{equation}
where 
\begin{equation}\label{TbarHW}
\bar{T}_c = z_0 T_c\;.
\end{equation}
In FIG. 2, we plot $\bar{T}_c$ as a function of $\bar{\mu} $ for $\omega l = 0$.  One notes that this hard wall result is in agreement with the QCD phase diagram, in the sense that the critical temperature decreases with the quark chemical potential. 

In order to compute the rotation effect on  the critical temperature, we must to obtain the critical horizons as functions of $\omega l$ and $\mu$, by replacing the dimensionless action density $\eqref{barDeltaE2}$ into the phase transition equation \eqref{HPtransition}. These horizons can be solved analytically, and the resulting critical temperatures also 
 appears
 as a function of the plasma rotational velocity and the chemical potential. In FIG. 3, using the real and positive critical horizon  solution in the analyzed region, we have plot $\bar{T}_c$ as a function $\bar{\mu}$ at fixed rotational velocities. One observes that rotation decreases the critical temperatures for each value of the chemical potential $\bar{\mu}$. 
 
This behavior is also illustrated in  FIG. 4, where we plot $\bar{T}_c$ as a function of $\omega l$ at fixed values of $\bar{\mu}$. The combination of the effects of density and rotation is shown in FIG. 5, which corresponds to the extended QCD phase diagram taking into account the plasma rotation.

\begin{figure}[!htb]
	\centering
	\includegraphics[scale=0.64]{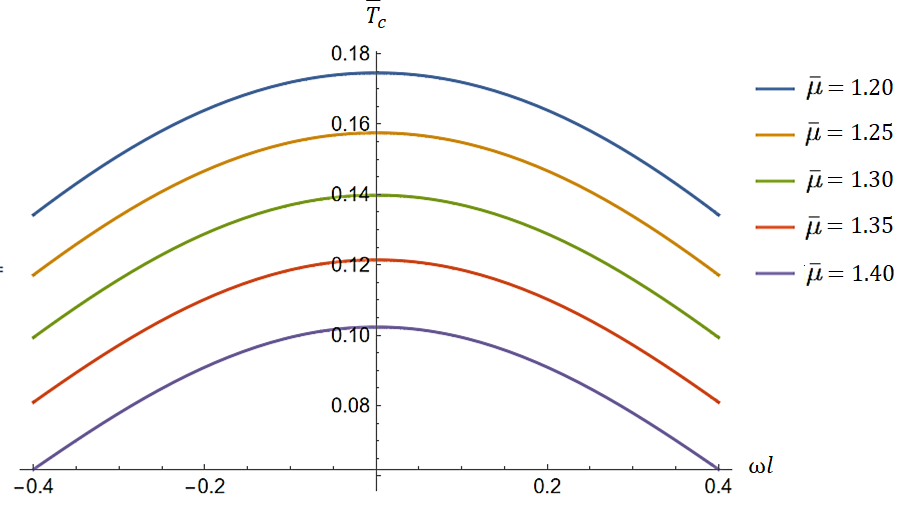}
	\caption{Critical temperatures versus QGP rotational velocity, for HP transitions at fixed quark chemical potentials $\bar{\mu}$ in the hard wall AdS/QCD model.  }
\end{figure}

\begin{figure}[!htb]
	\centering
	\includegraphics[scale=0.68]{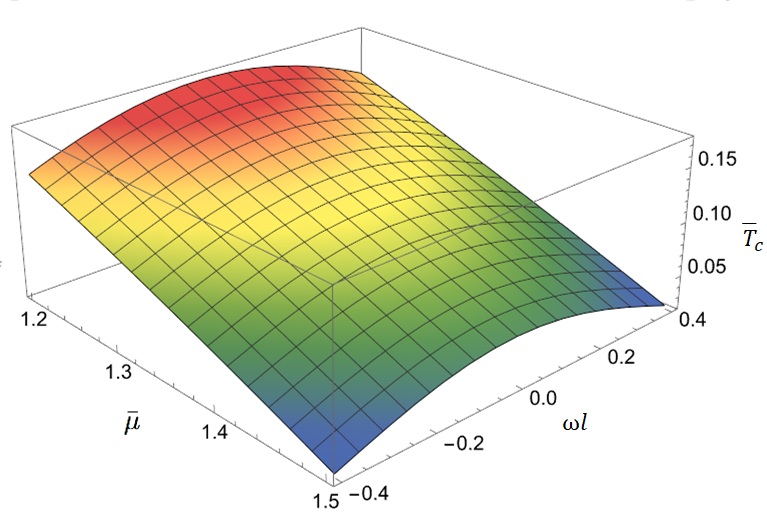}
	\caption{Extended QCD phase diagram. Critical temperature as a function of the quark chemical potential and plasma rotational velocity, $\bar{T}_c( \bar{\mu}, \omega l )$, in the hard wall AdS/QCD model. }
\end{figure}

For a non-rotating system with zero quark density, we recover the expression for the critical temperature obtained in \cite{Herzog:2006ra},  $T_c(0,0) = 2^{1/4}/(\pi z_0)$. In the hard wall model, the IR cutoff can be fixed by using QCD phenomenology, which yields $1/z_0 = 323 \,\text{MeV} $, according to the mass of the lightest $\rho$ meson, see again \cite{Herzog:2006ra}. This makes it possible to find the physical critical temperature
\begin{equation}
    T_c(0,0) = 122\, \text{MeV}\;.
\end{equation}
This value of $T_c(0,0)$ has a large discrepancy with respect to lattice QCD results. This situation improves in the AdS/QCD soft wall model. As we will see,   the   results of the phase transition in the presence of a scalar dilaton field will present a pattern similar to the ones found  in the hard wall model, but with higher critical temperatures.

\subsection{Soft wall model}
In the soft wall AdS/QCD model \cite{Karch:2006pv} one takes $ z_0 \to \infty $, while $c \neq 0$, in equations \eqref{DeltaEBH} and \eqref{DeltaEAdS}. In this case $z_{min} = z_h$ and the dimensionless regularized energy density is  
\begin{eqnarray}\label{deltaEsoft}
    \bigtriangleup \bar{\mathcal{E}}(\bar{\mu},\omega , \bar{z}_h) = \frac{e^{-\bar{z_h}^2}\pi \bar{z}_h \gamma(wl) } {  2 z_h^4\left(1-\frac{\bar{\mu}^2 \bar{z}_h^2}{2}\right) } \left[2(-1+\bar{z_h}^2) + e^{\bar{z}_h^2}(1+\bar{\mu}^2\bar{z}_h^2)  + 2\bar{z}_h^4e^{\bar{z_h}^2}\text{Ei}(-\bar{z_h}^2)-\bar{\mu}^2\bar{s}_1(\omega, \bar{z}_h) +\bar{\mu}^4\bar{s}_2(\omega, \bar{z}_h)\right]\;,\nonumber\\ 
\end{eqnarray}
where we have defined 
\begin{eqnarray}
    \bar{s}_1(\omega, \bar{z}_h) &=& \gamma^4\left[ 3 + 3l^4\omega^4 - \frac{6l^2 \omega^2}{\bar{z}_h^4}\left( 4 + 4\bar{z}_h^2 - \bar{z}_h^4\right)\right] \;,\\
    \bar{s}_2(\omega, \bar{z}_h) &=& \frac{12 l^2 \omega^2\gamma^4}{z_h^8}\left(6 + 4\bar{z}_h^2 + \bar{z}^4_h\right)\;,
\end{eqnarray}
and $\text{Ei}(x) = - \int_{-x}^\infty e^{-t}/t\,dt $. We also defined $\bigtriangleup \bar{\mathcal{E}} = \kappa^2 \bigtriangleup \mathcal{E}/(L^3 c^{3/2})$ and used the soft wall dimensionless variables:
\begin{eqnarray}
    \bar{z}_h &=& z_h \sqrt{c}\;,\nonumber\\
    \bar{\mu} &=& \mu/\sqrt{c}\;,\nonumber\\
    \bar{q} &=& q/c^{3/2}\;.\label{varsw}
    \end{eqnarray}
It is useful to define the dimensionless soft wall temperature:
\begin{equation}\label{TbarSW}
    \bar{T} = T/\sqrt{c} \,.
\end{equation}

As illustrated in FIG. 6, the critical horizon, where the regularized energy density vanishes, is affected by the quark chemical potential,  following the same behavior as in the hard wall. The transition equation \eqref{HPtransition} has no analytical solution in the soft wall. On TABLES 1 and 3, in the Appendix, we show the critical horizon at different values of $\bar{\mu}$, obtained using numerical methods. Analogously to the hard wall model, the values of $\bar{z}{_h{_c}}$  depend on the plasma rotation, appearing then as a function of $\bar{\mu}$ and $\omega$.  By comparing the soft and hard wall critical horizons at the same conditions, one concludes that the critical horizon position is slightly smaller in the soft wall, which leads to higher critical temperatures. 

\begin{figure}[!htb]
	\centering
	\includegraphics[scale=0.60]{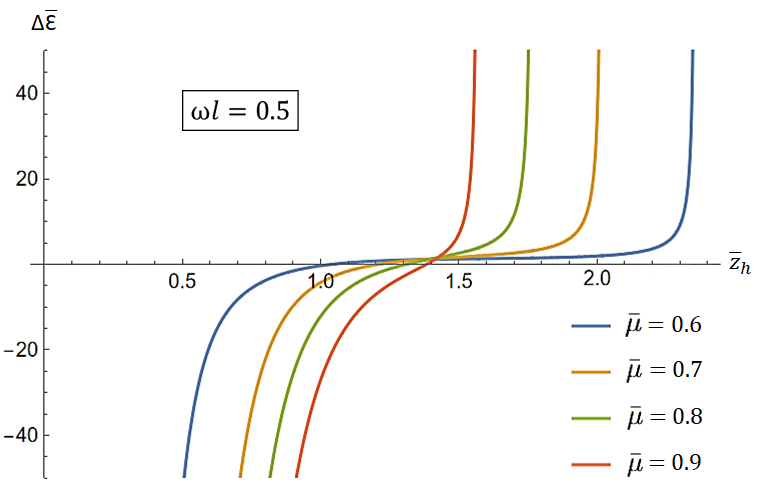}
	\caption{Action density of regularized rotating BH at finite density versus horizon position in the soft wall model, at a fixed rotational velocity ($\omega l = 0.5$), and different quark chemical potentials.}
\end{figure}

From the values of ${\bar z}_{h_c}$ of TABLES 1 and 3 in the appendix, we compute the critical temperatures of deconfinement in the soft wall, by replacing the critical quark chemical potential and its corresponding QGP rotational velocity into Eq. \eqref{HTrot}, assuming that $\eta = 1$ with $\bar{q}_c = \bar{\mu}_c/z^2{_h{_c}}$. The results are displayed in TABLES 2 and 4 in the Appendix. 

The soft wall non-rotating case was recently studied in \cite{Braga:2024nnj}. In order to analyze the rotation effect on $\bar{T}_c$, we plot in FIG. 7 the critical temperature versus $\bar{\mu}$, at different plasma rotational velocities. In turn, FIG. 8 shows a plot of $\bar{T}_c$ as a function of $\omega l$, at fixed values of $\bar{\mu}$. From these figures, one concludes that the critical temperature $\bar{T}_c(\bar{\mu},\omega l)$ of the soft wall model is given by a surface with a shape similar to the one obtained in the hard wall, see FIG. 5. This is an indication that this behavior does not depend on the way in which we introduce an IR energy scale. The result that $T_c$ decreases as $\omega l$ and/or $\mu$ increase is consistent with the one obtained from the analysis of Polyakov loops in the EMD model for a rotating QGP  \cite{Zhao:2022uxc}. 

\begin{figure}[!htb]
	\centering
	\includegraphics[scale=0.60]{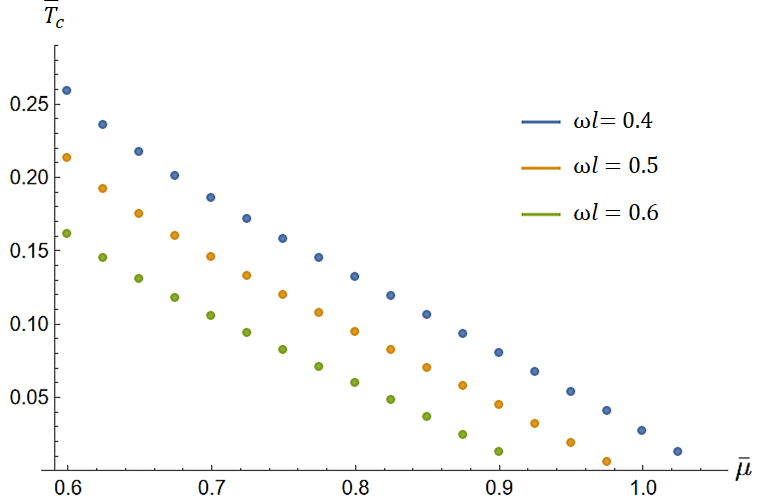}
	\caption{Critical temperatures of deconfinement as a function of quark chemical potential in the soft wall model, at fixed values of plasma rotational velocity.}
\end{figure}

\begin{figure}[!htb]
	\centering
	\includegraphics[scale=0.64]{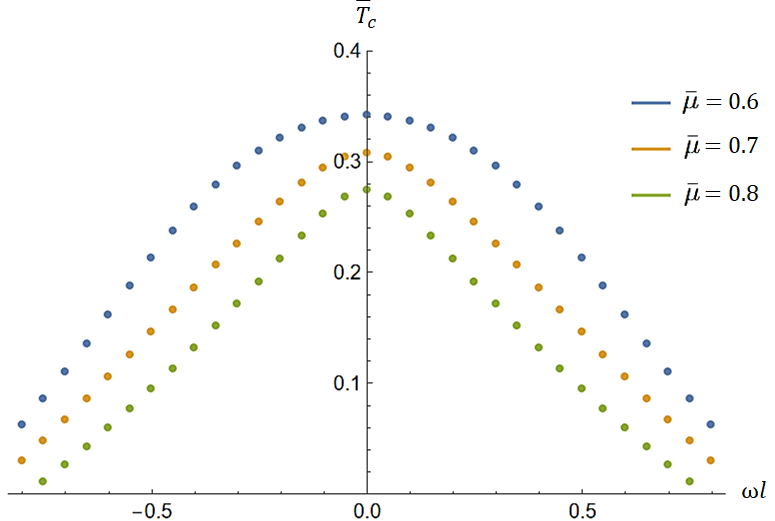}
	\caption{Critical temperatures versus QGP rotational velocities, for HP transitions at fixed quark chemical potentials in the soft wall AdS/QCD model.  }
\end{figure}

 One also observes that for both models there are critical value of $\bar{\mu}$,  that we call ${\bar \mu}_0$, for which the critical temperature vanishes: $T_c=0$. The values of ${\bar \mu}_0$ depend on the plasma rotation. 
 They correspond to the values of $\bar\mu$ where the curves in  Figures 3 and 7 touch the horizontal axis. Note that ${\bar \mu}_0$
 decreases as the QGP rotational velocity increases.

\section{Final remarks and conclusions}
 
Our results demonstrate that rotation  decreases the critical temperatures of deconfinement for a system at finite density. The critical quark chemical potential (${\bar\mu}_0$) for the transition to occur  at zero temperature is also affected by plasma rotation. In both models, the effect is the same: ${\bar\mu}_0$ tends to decrease as the rotation increases --- see FIG. 3 and FIG. 7 for the hard and soft wall models, respectively. 

In the hard wall, the critical temperature for the non-rotating case is given by expression \eqref{barTc} that can be written in the form
\begin{eqnarray}\label{Tcg1g2}
  T_c(\bar{\mu}) = T_c(0,0)\, g_1(\bar{\mu})\;,    \end{eqnarray}
where  
\begin{equation}
 g_1(\bar{\mu}) =   \left(-\bar{\mu}^2 + \frac{\sqrt{1 + 3 \bar{\mu}^2 + 2\bar{\mu}^4}}{\sqrt{2}}\right)^{-1/2} \left( 1 + \bar{\mu}^4 - \bar{\mu}^2 \frac{\sqrt{1 + 3 \bar{\mu}^2 + 2\bar{\mu}^4}}{2\sqrt{2}}  \right)\;.   
\end{equation}
The most critical chemical potential is defined by the phase transition at $\bar{T}_c = 0$. So that, for a non-rotating plasma, it is defined by
\begin{equation}
    g_1(\bar{\mu}_0) = 0\;, 
\end{equation}
which yields 
\begin{equation}\label{mu0w0HW}
    \bar{\mu}_0^{hw}(\omega l = 0) =     1.6429\;. 
\end{equation}
For  $\bar{\mu} \geq \bar{\mu}_0^{hw}$,   QCD matter is in the quark-gluon plasma phase, independent of the temperature. 
  
From FIG. 3, one concludes that $\bar{\mu}_0^{hw}$ is a function of the plasma rotational velocity, $\bar{\mu}_0^{hw} = \bar{\mu}_0^{hw}(\omega l)$. Using the values displayed in this figure, one obtains
\begin{eqnarray}
    \bar{\mu}_0^{hw} (0.25) &=& 1.6013\;,\nonumber\\
    \bar{\mu}_0^{hw} (0.30) &=& 1.5858\;,\nonumber\\
    \bar{\mu}_0^{hw} (0.35) &=& 1.5690\;,\nonumber\\
    \bar{\mu}_0^{hw} (0.40) &=& 1.5515\;,\nonumber\\
    \bar{\mu}_0^{hw} (0.45) &=& 1.5336\;,
\end{eqnarray}
that clearly show that $\bar{\mu}_0^{hw}$ decreases with $\omega l$, in accordance with the roots of $\bar{T}_c$ described by the curves plotted in FIG. 3.

For the soft wall model there is no analytical solution. However, from a numerical analysis one can obtain 
 the points shown in FIG. 7 and FIG. 8, which allows us to analyze the rotation effect on $\bar{\mu}_0$ in this model. For the case without rotation in this model, the numerical estimative is
\begin{equation}\label{mu0SW}
    \bar{\mu}_0^{sw}(\omega l = 0) \approx 1.4350\;.
\end{equation}
This result shows that $\bar{\mu}_0$ is sensitive to the way we introduce the IR parameter into the theory, see Eq. \eqref{mu0w0HW}. For the plasma rotational velocities used to plot FIG. 7, one obtains 
\begin{eqnarray}
    \bar{\mu}_0^{sw} (0.40) &\approx& 1.030\;,\nonumber\\
    \bar{\mu}_0^{sw} (0.50) &\approx& 0.975\;,\nonumber\\  
    \bar{\mu}_0^{sw} (0.60) &\approx& 0.905\;,
\end{eqnarray}
that indicates that the effect is the same as in the hard wall, in other words, the plasma rotation also decreases the value of $\mu_0$ in the soft wall model. 

Since $ \bar{\mu}_0^{hw}$ and $ \bar{\mu}_0^{sw}$ are written in terms of $z_0$ and $\sqrt{c}$, respectively, the physical value of $\mu_0$ depends on the IR parameter of each model. In order to recover the full dimensional values of the critical temperatures and densities, one needs to estimate the IR energy parameters. For the soft wall, a fit of the masses of lightest $\rho$-mesons \cite{Herzog:2006ra} leads to $\sqrt{c}=338\, \text{MeV}$. For the hard wall, as mentioned in Section 3.1,  QCD phenomenology yields $1/z_0 = 323 \,\text{MeV} $. The critical temperature of deconfinement in the soft wall for a non-rotating QCD matter with zero density is $T_c(0,0) = 0.491728 \sqrt{c} = 191\, \text{MeV}$, which is consistent with lattice QCD data, see \cite{Herzog:2006ra, Karsch:2006xs}. This contrasts with the hard wall model prediction of $ T_c(0,0) = 122 \, \text{MeV}$. For the most critical quark chemical potential, the holographic predictions of each model for the non-rotating plasma are 
\begin{eqnarray}
  \mu_0^{hw}(0) =  \mu_0^{hw}(0)/z_0 =   530.657 \, \text{MeV}\;, \\
  \mu_0^{sw}(0) =  \mu_0^{sw}(0)\sqrt{c} =   485.030 \, \text{MeV}\;.
\end{eqnarray}
The effect of plasma rotation tends to make these values increasingly smaller.

Moreover, for the system at finite density there is the parameter $\eta$ that  appears in the gauge field  solution \eqref{A0}. We have considered $\eta = 1 $ in the previous developments, just for simplification. However, $\eta$ affects the confinement/deconfinement transition in a non-trivial way \cite{Colangelo:2010pe}. The dependence on this parameter is recovered by defining the physical density  
\begin{eqnarray}
    \mu_{phys}(\eta) = \eta \mu\;, 
\end{eqnarray}
where $\mu$ is the quark chemical potential with $\eta = 1$. In these type of holographic models, being $N_c$ the rank of the gauge group, and $N_f$ the number of flavor, the BH charge and the parameter $q$ is related through, see \cite {Lee:2009bya},
\begin{eqnarray}
    Q = \sqrt{\frac{3N_c}{2N_f}} q\;.
\end{eqnarray}
By comparing the equation above with Eq. \eqref{eta}, one obtains
\begin{equation}
    \eta  = \sqrt{\frac{3N_c}{2N_f}}\;,
\end{equation}
which yields $\eta = \sqrt{3/4} $ for a system with 3 colors and 6 flavors. Therefore, the actual physical predictions for the critical quark chemical potential are the results presented in our previous calculations multiplied by this factor.    

 Besides that, considering the relation between the quark and baryon chemical potentials given by $\mu = \mu_B/3 $, see \cite{Colangelo:2010pe}, one obtains the following critical values of $\mu_B$ at $T=0$ for the non-rotating system in the hard and soft wall models, respectively,
\begin{eqnarray}
\mu_{0 B}^{hw}(0)  =  1378.687\,\text{MeV}\;,\\
\mu_{0 B}^{sw}(0) = 1260.144\,\text{MeV}\;.
\end{eqnarray}
Comparing these two results with the ones obtained from EMD models, see \cite{Chen:2020ath, Zhao:2022uxc},
one concludes that the predictions are of the same  order. See, for instance, the 3D-QCD phase diagram of FIG. 6 in \cite{Zhao:2022uxc}, where $\mu_{0 B}(0) \approx 1200 \, \text{MeV}$.

On the other hand, there is an important difference between the results obtained in \cite{Chen:2020ath, Zhao:2022uxc} and the ones found here. 
In \cite{Chen:2020ath, Zhao:2022uxc}, when the chemical potential is finite,  there is a crossover, in contrast to the first order transition between the confined and deconfined phases that was found in the present work for the analyzed values of $\mu $. The difference is associated to the fact that the models   and   the characterization of the transition are different.  
In \cite{Chen:2020ath, Zhao:2022uxc} there is just one  background geometry, obtained from an Einstein-Maxwell-dilaton holographic approach. The main confinement criterea is the behavior of a Polyakov loop, that represents the interaction energy of a static quark-antiquark pair. The geometry does not have any transition. It is the string that represents the loop that behaves differently in the two phases.

In contrast, here we represented the deconfinement process by a Hawking-Page phase transition between two different geometries. A thermal AdS one, representing the hadronic phase, and a black hole AdS space, representing the QGP phase.   
The transition is always of first order, as it happpens for the HP case. It is interesting to have alternative descriptions of such a non-trivial process as the formation of the QGP. More than that, it is clear that the deconfinement transition corresponds to a change in the medium. Or, to be more precise, to the formation of a thermal medium. Also, the geometric background is supposed to represent the medium. So, it is reasonable to suppose that the transition between the vacuum and the QGP should be holographically represented by a change in the dual geometries.

In Section 3.1, we found expression \eqref{FreeHW} for the   the regularized action density in the hard wall model. This expression corresponds to the free Gibbs energy, up to a $1/\beta$ factor, \textit{i. e.}, $\Phi_{Gibbs} = \bigtriangleup \mathcal{E} /\beta $. It emerges from this equation that the phase transition only occurs for $z_0 > z_h$, analogously to the non-rotating case at zero density. 
In FIG. 1, using the dimensionless variables of Eq. \eqref{dimensionlessvar}, we have plot  $\bigtriangleup \mathcal{E}$ as a function of $\bar{z}_h$ at finite density. In this case, $\bigtriangleup \mathcal{E}$ has a divergency at $\bar{z}_h = \sqrt{2}/\bar{\mu}$, which corresponds to the edge of the horizon domain, see Eq. \eqref{positivity}. As mentioned above, in FIG. 2 one finds a plot of $\bar{T}_c$ as a function of $\bar{\mu}$ at $\omega l = 0$, while FIG. 3 contains the curves of $\bar{T}_c$ at different rotational velocities. These figures show that the hard wall model is consistent with the QCD phase diagram for the rotating case: $T_c$ decreases as $\mu$ increases. On the other hand, in FIG. 3, the plot of $\bar{T}_c$ as a function of $\omega l$, at fixed chemical potentials, shows that $T_c$ also decreases with the rotational velocity. This result is a generalization of the one obtained in \cite{Braga:2022yfe} for a rotating plasma with zero density. The combination of both behaviors leads to the 3D QCD phase diagram of rotating QCD matter of FIG 5. Such a result is consistent with the diagram generated from the analysis of Polyakov loops in the holographic EMD model, see \cite{Zhao:2022uxc}.

Concerning the soft wall model, we would like to emphasize that the values of $z_h$ in the phase transitions legitimize the approximation used in this work, where we have assumed the RN solution in the presence of the dilaton, see Eq. \eqref{V}. The RN solution is exact for the hard wall, but only valid for small values of $\bar{z}$ in the soft wall. In both models, along the Hawking-Page curves, the dimensionless critical horizon positions are small --- see TABLES 1 and 3 for the soft wall model. The fact that $z_h$ is small at high densities and low temperatures is explained by Eq. \eqref{positivity}, that shows that $z_h$ is limited at finite density, with its upper limit being inversely proportional to the BH charge (or to the quark chemical potential). The exact solution for the the vector field in the soft wall can be found in \cite{Andreev:2010bv}. Naturally, in the limit of small $\bar{z}$, it recovers the RN approximation.

For the soft wall case, discussed in Section 3.2, the HP phase transition equation \eqref{HPtransition} has no analytical solution. In FIG. 6, one finds a plot of $\bigtriangleup \bar{\mathcal{E}}$ at fixed chemical potentials, that shows the same behavior of the hard wall one. The critical horizon in the soft wall is slightly lower than the hard wall one. It is observed that 
$\bigtriangleup \bar{\mathcal{E}}$ of the soft wall also has a divergency at  $\bar{z}_h = \sqrt{2}/\bar{\mu}$.   From a numerical analysis, the critical horizons at different quark chemical potentials were calculated and shown in TABLE 1. The critical horizons depends on plasma rotation and increases with $\bar{\mu}$,  the same behavior was found in the hard wall.

The numerical results for the critical temperatures in the soft wall model are displayed in TABLE 2 and TABLE 4 of the Appendix and ploted  in FIG. 7 and FIG. 8, respectively. Comparing   with FIG. 2 and FIG. 3, one concludes that the two models have similar behaviors: $\bar{T}_c$ decreases when $\bar{\mu}$ and/or $\omega l $ increase.

The numerical differences between the two models can be associated to the difference in the way that the infrared (IR) energy parameter is introduced. The hard wall is a much simpler model, in terms of calculations. However, the IR scale is introduced in a non-analytic way, by just cutting abruptly the space at some value of the coordinate $z$. More than that, the IR scale is not present in the Einstein gravity Lagrangean but only in the integration limit. In contrast, in the soft wall case there is no problem of analyticity, and the IR parameter affects the whole space. In this sense the results of the SW model should be taken in general as more reliable, although the simpler results of the HW model show the qualitative aspects of the phase transitions in a similar way. On the other hand, as mentioned before, the RN solution for the gauge field of Eq. \eqref{V} is exact for the hard wall, but only approximate for small values of $\bar{z}$ in the soft wall.

In conclusion, both models analyzed in this work predict that the effects of rotation and density in the QGP are not independent. The plasma rotation contributes to a decrease in the critical temperature of dissociation and/or in the critical quark chemical potential. These results are in agreement with experimental data, which indicates that QCD phase diagram 
has a critical density that is sensitive to quark-gluon plasma vorticity.

\noindent \textbf{Acknowledgments}: The authors are supported by FAPERJ --- Fundação Carlos Chagas Filho de Amparo à Pesquisa do Estado do Rio de Janeiro, CNPq - Conselho Nacional de Desenvolvimento Cient\'ifico e Tecnol\'ogico and Coordena\c c\~ao de Aperfei\c coamento de Pessoal de N\'ivel Superior - Brasil (CAPES) - Finance Code 001. O. C. Junqueira also thanks to the S\~ao Paulo Research Foundation (FAPESP) Grant No. 2024/14390-0 for partial financial support.

\appendix

\section{Auxiliary tables with numerical results}

\begin{table}[h]
\centering
\begin{tabular}[c]{|c|c|c|c|}
\hline 
\text{}  & $\,\,\,\omega l=0.4\,\,\, $ & $\,\,\,\omega l=0.5\,\,\, $ & $\,\,\,\omega l=0.6\,\,\, $\\
 \hline
 $\,\,\,\bar{\mu}=0.600\,\,\, $ &$\,\,\,0.945817\,\,\,$ &$\,\,\,1.041991\,\,\,$&$\,\,\,1.182244\,\,\,$  \\
\hline
 $\,\,\,\bar{\mu}=0.625\,\,\,$ & $\,\,\, 0.996634\,\,\,$ &$\,\,\,1.098487\,\,\,$&$\,\,\,1.235145\,\,\,$  \\
\hline
 $\,\,\,\bar{\mu}=0.650\,\,\,$ & $\,\,\,1.037177\,\,\,$ &$\,\,\,1.142511\,\,\,$&$\,\,\,1.277314\,\,\,$  \\
\hline 
 $\,\,\,\bar{\mu}=0.675\,\,\,$ & $\,\,\,1.071658\,\,\,$  &$\,\,\,1.179399\,\,\,$&$\,\,\,1.312959\,\,\,$  \\
\hline
 $\,\,\,\bar{\mu}=0.700\,\,\, $ & $\,\,\,1.101974\,\,\,$ &$\,\,\,1.211481\,\,\,$&$\,\,\,1.344075\,\,\,$  \\
\hline
 $\,\,\,\bar{\mu}=0.725\,\,\,$ &$\,\,\,1.129173\,\,\,$ &$\,\,\,1.240018\,\,\,$&$\,\,\, 1.371802\,\,\,$  \\ 
\hline
 $\,\,\,\bar{\mu}=0.750\,\,\,$ &$\,\,\,1.153912\,\,\,$ &$\,\,\,1.265794\,\,\,$&$\,\,\,1.396865\,\,\,$   \\
\hline 
$\,\,\,\bar{\mu}=0.775\,\,\, $ &$\,\,\,1.176635\,\,\,$  &$\,\,\,1.289335\,\,\,$&$\,\,\,1.419762\,\,\,$ \\
\hline
$\,\,\,\bar{\mu}=0.800\,\,\,$ &$\,\,\, 1.197665\,\,\,$  &$\,\,\,1.311019\,\,\,$&$\,\,\,1.440856\,\,\,$\\
\hline
 $\,\,\,\bar{\mu}=0.825\,\,\,$ &$\,\,\,1.217244\,\,\,$  &$\,\,\,1.331127\,\,\,$&$\,\,\,1.460419\,\,\,$ \\ 
\hline
$\,\,\,\bar{\mu}=0.850\,\,\,$ &$\,\,\,1.235565\,\,\,$  &$\,\,\,1.349878\,\,\,$&$\,\,\,1.478665\,\,\,$  \\
\hline  $\,\,\,\bar{\mu}=0.875\,\,\,$&$\,\,\,1.252779\,\,\,$  &$\,\,\,1.367446\,\,\,$&$\,\,\,1.495763\,\,\,$  \\
\hline
 $\,\,\,\bar{\mu}=0.900\,\,\,$ &$\,\,\,1.269013\,\,\,$  &$\,\,\,1.383973\,\,\,$&$\,\,\,1.511851\,\,\,$ \\ 
\hline
 $\,\,\,\bar{\mu}=0.925\,\,\,$ &$\,\,\,1.284370\,\,\,$  &$\,\,\,1.399576\,\,\,$&$\,\,\,\textbf{x}\,\,\,$  \\
\hline 
 $\,\,\,\bar{\mu}=0.950\,\,\,$ &$\,\,\,1.298939\,\,\,$ &$\,\,\,1.414353\,\,\,$&$\,\,\,\textbf{x}\,\,\,$  \\
\hline
 $\,\,\,\bar{\mu}=0.975\,\,\,$ &$\,\,\, 1.312797\,\,\,$ &$\,\,\,1.428387\,\,\,$&$\,\,\,\textbf{x}\,\,\,$ \\
\hline
$\,\,\,\bar{\mu}=1.000\,\,\,$ &$\,\,\,1.326007\,\,\,$   &$\,\,\,\textbf{x}\,\,\,$&$\,\,\,\textbf{x}\,\,\,$ \\
\hline 
 $\,\,\,\bar{\mu}=1.025\,\,\,$ &$\,\,\,1.338629\,\,\,$ &$\,\,\,\textbf{x}\,\,\,$&$\,\,\,\textbf{x}\,\,\,$  \\
\hline
 \end{tabular}   
\caption{Critical horizon positions of confinement/deconfinement phase transition in the soft wall AdS/QCD model, at different chemical potentials ($\bar{\mu}$) and rotational velocities ($\omega l$), used to compute the critical temperatures of TABLE 2. (The ``$\textbf{x}$" represents the temperatures that do not obey the physical positivity condition of Eq. \eqref{positivity}, and were not used in FIG. 7.) }
\label{table1}
\end{table}

\begin{table}[h]
\centering
\begin{tabular}[c]{|c|c|c|c|}
\hline 
\text{}  & $\,\,\,\omega l=0.4\,\,\, $ & $\,\,\,\omega l=0.5\,\,\, $ & $\,\,\,\omega l=0.6\,\,\, $\\
 \hline
 $\,\,\,\bar{\mu}=0.600\,\,\, $ &$\,\,\,0.258781\,\,\,$ &$\,\,\,0.212852\,\,\,$&$\,\,\,0.161204\,\,\,$  \\
\hline
 $\,\,\,\bar{\mu}=0.625\,\,\,$ & $\,\,\, 0.235933\,\,\,$ &$\,\,\,0.191806\,\,\,$&$\,\,\,0.144737\,\,\,$  \\
\hline
 $\,\,\,\bar{\mu}=0.650\,\,\,$ & $\,\,\,0.174746\,\,\,$ &$\,\,\,0.174746\,\,\,$&$\,\,\,0.130650\,\,\,$  \\
\hline 
 $\,\,\,\bar{\mu}=0.675\,\,\,$ & $\,\,\,0.201005\,\,\,$  &$\,\,\,0.159667\,\,\,$&$\,\,\,0.117782\,\,\,$  \\
\hline
 $\,\,\,\bar{\mu}=0.700\,\,\, $ & $\,\,\,0.185975\,\,\,$ &$\,\,\,0.145723\,\,\,$&$\,\,\,0.105604\,\,\,$  \\
\hline
 $\,\,\,\bar{\mu}=0.725\,\,\,$ &$\,\,\,0.171786\,\,\,$ &$\,\,\,0.132470\,\,\,$&$\,\,\, 0.093823\,\,\,$  \\ 
\hline
 $\,\,\,\bar{\mu}=0.750\,\,\,$ &$\,\,\,0.158144\,\,\,$ &$\,\,\,0.119642\,\,\,$&$\,\,\,0.082257\,\,\,$   \\
\hline 
$\,\,\,\bar{\mu}=0.775\,\,\, $ &$\,\,\,0.144854\,\,\,$  &$\,\,\,0.107065\,\,\,$&$\,\,\,0.070785\,\,\,$ \\
\hline
$\,\,\,\bar{\mu}=0.800\,\,\,$ &$\,\,\, 0.131779\,\,\,$  &$\,\,\,0.094619\,\,\,$&$\,\,\,0.059322\,\,\,$\\
\hline
 $\,\,\,\bar{\mu}=0.825\,\,\,$ &$\,\,\,0.118819\,\,\,$  &$\,\,\,0.082215\,\,\,$&$\,\,\,0.047806\,\,\,$ \\ 
\hline
$\,\,\,\bar{\mu}=0.850\,\,\,$ &$\,\,\,0.105900\,\,\,$  &$\,\,\,0.069788\,\,\,$&$\,\,\,0.036190\,\,\,$  \\
\hline  $\,\,\,\bar{\mu}=0.875\,\,\,$&$\,\,\,0.092960\,\,\,$  &$\,\,\,0.057287\,\,\,$&$\,\,\,0.024435\,\,\,$  \\
\hline
 $\,\,\,\bar{\mu}=0.900\,\,\,$ &$\,\,\,0.079954\,\,\,$  &$\,\,\,0.044671\,\,\,$&$\,\,\,0.012514\,\,\,$ \\ 
\hline
 $\,\,\,\bar{\mu}=0.925\,\,\,$ &$\,\,\,0.066843\,\,\,$  &$\,\,\,0.031907\,\,\,$&$\,\,\,\textbf{x}\,\,\,$  \\
\hline 
 $\,\,\,\bar{\mu}=0.950\,\,\,$ &$\,\,\,0.053596\,\,\,$ &$\,\,\,0.018969\,\,\,$&$\,\,\,\textbf{x}\,\,\,$  \\
\hline
 $\,\,\,\bar{\mu}=0.975\,\,\,$ &$\,\,\, 0.040185\,\,\,$ &$\,\,\,0.005833\,\,\,$&$\,\,\,\textbf{x}\,\,\,$ \\
\hline
$\,\,\,\bar{\mu}=1.000\,\,\,$ &$\,\,\,0.026589\,\,\,$   &$\,\,\,\textbf{x}\,\,\,$&$\,\,\,\textbf{x}\,\,\,$ \\
\hline 
 $\,\,\,\bar{\mu}=1.025\,\,\,$ &$\,\,\,0.012788\,\,\,$ &$\,\,\,\textbf{x}\,\,\,$&$\,\,\,\textbf{x}\,\,\,$  \\
\hline
 \end{tabular}   
\caption{Critical temperatures of deconfinement in the soft wall model at different chemical potentials and rotational velocities, corresponding to the points of FIG. 7.}
\label{table2}
\end{table} 

\begin{table}[h]
\centering
\begin{tabular}[c]{|c|c|c|c|}
\hline 
\text{}  & $\,\,\,\bar{\mu}=0.6\,\,\, $ & $\,\,\,\bar{\mu}=0.7\,\,\, $ & $\,\,\,\bar{\mu}=0.8\,\,\, $\\
 \hline
 $\,\,\,\omega l=0.00\,\,\, $ &$\,\,\,0.820006\,\,\,$ &$\,\,\,0.850888\,\,\,$&$\,\,\,0.877052\,\,\,$  \\
\hline
 $\,\,\,\omega l=0.05\,\,\,$ & $\,\,\, 0.821576\,\,\,$ &$\,\,\,0.857176\,\,\,$&$\,\,\,0.888259\,\,\,$  \\
\hline
 $\,\,\,\omega l=0.10\,\,\,$ & $\,\,\,0.826351\,\,\,$ &$\,\,\,0.874735\,\,\,$&$\,\,\,0.917275\,\,\,$  \\
\hline 
 $\,\,\,\omega l=0.15\,\,\,$ & $\,\,\,0.834534\,\,\,$  &$\,\,\,0.900799\,\,\,$&$\,\,\,0.956235\,\,\,$  \\
\hline
 $\,\,\,\omega l=0.20\,\,\, $ & $\,\,\,0.846481\,\,\,$ &$\,\,\,0.932896\,\,\,$&$\,\,\,1.000085\,\,\,$  \\
\hline
 $\,\,\,\omega l=0.25\,\,\,$ &$\,\,\,0.862725\,\,\,$ &$\,\,\,0.969503\,\,\,$&$\,\,\, 1.046587\,\,\,$  \\ 
\hline
 $\,\,\,\omega l=0.30\,\,\,$ &$\,\,\,0.884008\,\,\,$ &$\,\,\,1.009922\,\,\,$&$\,\,\,1.094993\,\,\,$   \\
\hline 
$\,\,\,\omega l=0.35\,\,\, $ &$\,\,\,0.911303\,\,\,$  &$\,\,\,1.054010\,\,\,$&$\,\,\,1.145255\,\,\,$ \\
\hline
$\,\,\,\omega l=0.40\,\,\,$ &$\,\,\, 0.945816\,\,\,$  &$\,\,\,1.101974\,\,\,$&$\,\,\,1.197664\,\,\,$\\
\hline
 $\,\,\,\omega l=0.45\,\,\,$ &$\,\,\,0.988920\,\,\,$  &$\,\,\,1.154257\,\,\,$&$\,\,\,1.252706\,\,\,$ \\ 
\hline
$\,\,\,\omega l=0.50\,\,\,$ &$\,\,\,1.041990\,\,\,$  &$\,\,\,1.211481\,\,\,$&$\,\,\,1.311018\,\,\,$  \\
\hline  $\,\,\,\omega l=0.55\,\,\,$&$\,\,\,1.106181\,\,\,$  &$\,\,\,1.274428\,\,\,$&$\,\,\,1.373398\,\,\,$  \\
\hline
 $\,\,\,\omega l=0.60\,\,\,$ &$\,\,\,1.182243\,\,\,$  &$\,\,\,1.344075\,\,\,$&$\,\,\,1.440856\,\,\,$ \\ 
\hline
 $\,\,\,\omega l=0.65\,\,\,$ &$\,\,\,1.270594\,\,\,$  &$\,\,\,1.421679\,\,\,$&$\,\,\,1.514722\,\,\,$  \\
\hline 
 $\,\,\,\omega l=0.70\,\,\,$ &$\,\,\,1.371742\,\,\,$ &$\,\,\,1.508975\,\,\,$&$\,\,\,1.596849\,\,\,$  \\
\hline
 $\,\,\,\omega l=0.75\,\,\,$ &$\,\,\, 1.487032\,\,\,$ &$\,\,\,1.608577\,\,\,$&$\,\,\,1.690005\,\,\,$ \\
\hline
$\,\,\,\omega l=0.80\,\,\,$ &$\,\,\,1.619792\,\,\,$   &$\,\,\,1.724834\,\,\,$&$\,\,\,\textbf{x}\,\,\,$ \\
\hline 
 \end{tabular}   
\caption{Critical horizon positions in the soft wall AdS/QCD model, used to compute the critical temperatures of TABLE 4. }
\label{table3}
\end{table} 

\begin{table}[h]
\centering
\begin{tabular}[c]{|c|c|c|c|}
\hline 
\text{}  & $\,\,\,\bar{\mu}=0.6\,\,\, $ & $\,\,\,\bar{\mu}=0.7\,\,\, $ & $\,\,\,\bar{\mu}=0.8\,\,\, $\\
 \hline
 $\,\,\,\omega l=0.00\,\,\, $ &$\,\,\,0.341197\,\,\,$ &$\,\,\,0.307734\,\,\,$&$\,\,\,0.273596\,\,\,$  \\
\hline
 $\,\,\,\omega l=0.05\,\,\,$ & $\,\,\, 0.339940\,\,\,$ &$\,\,\,0.304119\,\,\,$&$\,\,\,0.267540\,\,\,$  \\
\hline
 $\,\,\,\omega l=0.10\,\,\,$ & $\,\,\,0.336160\,\,\,$ &$\,\,\,0.294194\,\,\,$&$\,\,\,0.252313\,\,\,$  \\
\hline 
 $\,\,\,\omega l=0.15\,\,\,$ & $\,\,\,0.329833\,\,\,$  &$\,\,\,0.279911\,\,\,$&$\,\,\,0.232813\,\,\,$  \\
\hline
 $\,\,\,\omega l=0.20\,\,\, $ & $\,\,\,0.320922\,\,\,$ &$\,\,\,0.263029\,\,\,$&$\,\,\,0.212043\,\,\,$  \\
\hline
 $\,\,\,\omega l=0.25\,\,\,$ &$\,\,\,0.309382\,\,\,$ &$\,\,\,0.244690\,\,\,$&$\,\,\, 0.191264\,\,\,$  \\ 
\hline
 $\,\,\,\omega l=0.30\,\,\,$ &$\,\,\,0.295174\,\,\,$ &$\,\,\,0.225533\,\,\,$&$\,\,\,0.170908\,\,\,$   \\
\hline 
$\,\,\,\omega l=0.35\,\,\, $ &$\,\,\,0.278287\,\,\,$  &$\,\,\,0.205898\,\,\,$&$\,\,\,0.151082\,\,\,$ \\
\hline
$\,\,\,\omega l=0.40\,\,\,$ &$\,\,\, 0.258781\,\,\,$  &$\,\,\,0.185975\,\,\,$&$\,\,\,0.131779\,\,\,$\\
\hline
 $\,\,\,\omega l=0.45\,\,\,$ &$\,\,\,0.236845\,\,\,$  &$\,\,\,0.165884\,\,\,$&$\,\,\,0.112966\,\,\,$ \\ 
\hline
$\,\,\,\omega l=0.50\,\,\,$ &$\,\,\,0.212852\,\,\,$  &$\,\,\,0.145723\,\,\,$&$\,\,\,0.094619\,\,\,$  \\
\hline  $\,\,\,\omega l=0.55\,\,\,$&$\,\,\,0.187391\,\,\,$  &$\,\,\,0.125592\,\,\,$&$\,\,\,0.076730\,\,\,$  \\
\hline
 $\,\,\,\omega l=0.60\,\,\,$ &$\,\,\,0.161204\,\,\,$  &$\,\,\,0.105604\,\,\,$&$\,\,\,0.059322\,\,\,$ \\ 
\hline
 $\,\,\,\omega l=0.65\,\,\,$ &$\,\,\,0.135056\,\,\,$  &$\,\,\,0.085892\,\,\,$&$\,\,\,0.042447\,\,\,$  \\
\hline 
 $\,\,\,\omega l=0.70\,\,\,$ &$\,\,\,0.109587\,\,\,$ &$\,\,\,0.066605\,\,\,$&$\,\,\,0.026196\,\,\,$  \\
\hline
 $\,\,\,\omega l=0.75\,\,\,$ &$\,\,\, 0.085231\,\,\,$ &$\,\,\,0.047912\,\,\,$&$\,\,\,0.010719\,\,\,$ \\
\hline
$\,\,\,\omega l=0.80\,\,\,$ &$\,\,\,0.062223\,\,\,$   &$\,\,\,0.030019\,\,\,$&$\,\,\,\textbf{x}\,\,\,$ \\
\hline 
 \end{tabular}   
\caption{Critical temperatures  in the soft wall model, corresponding to the points of FIG. 8. }
\label{table4}
\end{table}

\clearpage

\bibliographystyle{utphys2}
\bibliography{library}

%\bibliographystyle{ieeetr}
%\bibliography{library}
\end{document}